\documentclass[twocolumn]{aastex631}

\usepackage{siunitx}
\usepackage{amsmath}
\usepackage{natbib}
\usepackage[normalem]{ulem}
\hypersetup{}
\usepackage{diagbox}

\newcommand{\eo}{\textcolor{cyan}{}\bgroup\markoverwith{\textcolor{red}{\rule[.5ex]{2pt}{2.5pt}}}\ULon}

\usepackage{ragged2e}

\definecolor{firebrick}{rgb}{0.7, 0.13, 0.13}

\begin{document}

\title{Galactic-scale Quasar Pairs from the Sloan Digital Sky Survey and Gaia DR3}

\email{yuanzhe@illinois.edu}
\author[0009-0001-5163-5781]{Yuanzhe Jiang}
\affiliation{Department of Astronomy, University of Illinois at Urbana-Champaign, Urbana, IL 61801, USA}

\author[0000-0003-1659-7035]{Yue Shen}
\affiliation{Department of Astronomy, University of Illinois at Urbana-Champaign, Urbana, IL 61801, USA}
\affiliation{National Center for Supercomputing Applications, University of Illinois at Urbana-Champaign, Urbana, IL 61801, USA}

\author[0000-0003-0049-5210]{Xin Liu}
\affiliation{Department of Astronomy, University of Illinois at Urbana-Champaign, Urbana, IL 61801, USA}
\affiliation{National Center for Supercomputing Applications, University of Illinois at Urbana-Champaign, Urbana, IL 61801, USA}

\author[0000-0001-6100-6869]{Nadia Zakamska}
\affiliation{Department of Physics and Astronomy, Bloomberg Center, Johns Hopkins University, Baltimore, MD 21218, USA}

\author[0000-0001-5105-2837]{Ming-Yang Zhuang}
\affiliation{Department of Astronomy, University of Illinois at Urbana-Champaign, Urbana, IL 61801, USA}

\author[0000-0001-7681-9213]{Arran C. Gross}
\affiliation{Department of Astronomy, University of Illinois at Urbana-Champaign, Urbana, IL 61801, USA}

\author[0000-0002-1605-915X]{Junyao Li}
\affiliation{Department of Astronomy, University of Illinois at Urbana-Champaign, Urbana, IL 61801, USA}

\author[0000-0002-9932-1298]{Yu-Ching Chen}
\affiliation{Department of Physics and Astronomy, Bloomberg Center, Johns Hopkins University, Baltimore, MD 21218, USA}

\author[0000-0001-7572-5231]{Yuzo Ishikawa}
\affiliation{MIT Kavli Institute for Astrophysics and Space Research, Massachusetts Institute of Technology, 
Cambridge, MA 02139}

\begin{abstract}
We preform a systematic search for galactic-scale quasar pairs and small-scale ($<3\arcsec$) lenses using the SDSS DR16 quasar catalog and Gaia DR3. Candidate double quasars (both are unobscured) are identified as Gaia resolved pairs around spectroscopically confirmed SDSS quasars ($L_{\rm bol} > 10^{44.5} \ {\rm erg \ s^{-1}}$) at $0.5 < z \lesssim 4.5$. Gaia astrometric information and SDSS spectral decomposition are used to exclude foreground star superpositions, which dominate ($\gtrsim 80\%$ of) the pair sample. We identify $136$ double quasar candidates from $1120$ Gaia-resolved pairs after a magnitude and redshift cut of $G<20.5$ and $z>0.5$ (803 double quasars out of 2,497 pairs without any cuts applied) with separations of $\sim 0\farcs3 - 3\arcsec$, corresponding to projected physical separations of $\sim 3 - 30$ kpc at the median redshift of the sample of $z = 1.7$.  We estimate an overall double quasar (lens and physical pairs combined) fraction using this sample, corrected for pair-resolving completeness, of $5.7_{-0.3}^{+0.3} \times 10^{-4}$ (bootstrapping errors). This double quasar fraction increases toward smaller separations, consistent with earlier findings. We also find little redshift evolution of the double quasar fraction for the luminous SDSS sample, consistent with previous observations and simulation predictions. However, the observed fraction is lower than simulation predictions {by $\sim$0.8 - 1.6 dex}, suggesting a significant population of obscured quasar pairs are missed in our search. Future wide-area space missions targeting both unobscured and obscured quasar pairs at sub-arcsec resolutions will reveal this population of obscured quasar pairs, and extend to much lower AGN luminosities.  
\end{abstract}


\section{Introduction}
\label{sec:intro}

The statistics of galactic-scale, i.e., with projected pair separations of $\lesssim 30\,{\rm kpc}$, dual supermassive black holes (SMBHs) critically constrain the dynamical evolution of pairs of SMBHs following the merger of their host galaxies. This is the progenitor population of gravitationally bound SMBHs at $\lesssim$ few parsec scales, whose eventual coalescence produces the loudest low-frequency gravitational waves (GW) to be detected with future GW facilities \citep[e.g.,][]{Kelly2017,Pau2023}. Over the past few years, there has been significant progress in both theoretical and observational studies of galactic-scale dual SMBHs. On the theoretical side, state-of-the-art hydrodynamic simulations with large volume are starting to produce sufficient statistics to predict the abundance and evolution of dual SMBHs and active galactic nuclei (AGNs) over cosmic time \citep[e.g.,][]{Chen2023,LiKy2023,Tiziana2023,Saeedzadeh_2024,Clara2025}. On the observational side, large-area sky surveys with imaging and spectroscopy are producing a large number of candidate dual AGNs to be compared with simulation predictions \citep[e.g.,][]{Sandoval_2023,Pfeifle_2024,Li2024}.

While a full comparison between observations and simulations is still lacking, recent studies have shown promise in reaching a consensus on the evolution of the dual SMBH population: (i) both observations and simulations suggest that the fraction of dual AGNs among the parent AGN population increases with decreasing AGN luminosity \citep[e.g.,][]{Chen2023}; (ii) a significant fraction of dual AGNs in simulations are obscured, and often missed from observational searches targeting unobscured broad-line AGNs \citep{Chen2023}; (iii) the dual AGN fraction seems to be elevated at the few-kpc scales compared with the fraction at larger separations \citep{Shen2023}; (iv) there is marginal evidence for redshift evolution of the dual AGN fraction \citep[e.g.,][]{Li2024}.  

In this work, we present an observational search for luminous dual AGNs, or dual quasars (the most luminous subset of AGNs), by combining SDSS and Gaia data. The SDSS DR16 quasar catalog \citep{Lyke2020} provides the input quasar sample, and the Gaia DR3 catalog \citep[DR3;][]{Babusiaux2023} provides resolved pairs around these luminous AGNs (or quasars) down to sub-arcsec separations. This work is an extension of an earlier work \citep{Shen2023} that focused on the $z>1.5$ regime. Our new sample extends to lower redshifts at $z>0.5$, as well as slightly fainter fluxes of $G<20.5$, compared with the threshold of $G<20.25$ used in \citet{Shen2023}. This expanded sample and better statistics will enable an investigation on the redshift evolution of the dual quasar fraction, as well as the dual quasar fraction as a function of pair separation. 

The Gaia DR3 catalog have precise coordinates, magnitudes, and astrometric measurements for all-sky sources to as faint as $G \sim 21$, which provides several advantages to search for quasar pairs. First, the nominal $\sim 0\farcs2$ resolution can resolve extremely close companions around distant quasars. Second, as demonstrated in \citet{Shen2023}, the accurate Gaia proper-motion measurements enable an efficient method to separate stars and quasars, a unique advantage over previous quasar pair searches based on ground-based photometric color selection.

In this work, we focus on the luminous unobscured (broad-line) quasars exclusively, given the nature and survey depths of SDSS and Gaia. By default, ``quasar pairs'' or ``dual quasars'' \citep[e.g.,][]{Comerford2009} refers to physically-associated quasar pairs inside the merging galaxies, instead of unrelated, projected quasar pairs at different redshifts. A common contaminant to dual quasar searches is gravitationally lensed quasars, which are often difficult to exclude from the sample without detailed follow-up observations and/or modelling \citep[e.g.,][]{Chen2022d,Gross2024,Ji_etal_2024}. For convenience, we use the term ``double quasars'' to collectively describe both quasar pairs and lensed quasars. We adopt a flat $\rm \Lambda CDM$ cosmology with $\rm \Omega_{\Lambda} = 0.7$, $\Omega_M = 0.3$, and  $H_0 = 70 \rm \ km \ s^{-1} \ Mpc^{-1}$. Pair physical separations are measured in proper units, and correspond to the projected separations.

\section{Data}
\label{sec:data}

\subsection{Crossmatch and Classification} 
\label{sec:match}

We start from the Sloan Digital Sky Survey Data Release 16 quasar catalog \citep[DR16Q;][]{Lyke2020} with improved systemic redshifts from \citet{Wu2022}, which includes 750,414 spectroscopically confirmed quasars and their properties. First, we search for Gaia DR3 sources in a $3\arcsec$-radius circular region around each SDSS quasar and identify 492,724 systems with at least one Gaia sources near the position of the DR16Q quasar. Given the depths of the SDSS DR16Q and Gaia, not all SDSS quasars are detected in Gaia. Among these Gaia-matched quasars, 2,524 systems have two Gaia detections around the DR16 quasar. 27 pairs have both components classified as bona fide quasars in DR16Q, resulting in double counting. After removing these duplicates, our initial Gaia-resolved pair sample includes 2,497 unique pairs. The detailed descriptions of crossmatch samples can be found in Table \ref{tbl:sample}. 

Similar to \cite{Shen2023}, we focus on Gaia-resolved double sources within $3\arcsec$ of the SDSS quasar position. The completeness of systems with more than two Gaia sources within this radius is significantly lower and hard to quantify. Matched multiple systems constitute only $\sim 0.7\%$ of double systems \citep{Shen2023} and are therefore negligible for subsequent statistical analyses. As a result, we ignore these higher-order multiple populations. Additionally, quasars with only one matched Gaia source may still contain subarcsecond quasar pairs, which require other methods and additional Gaia parameters to identify \citep[e.g.,][]{Shen2019, Hwang2020, Chen2022d, Makarov2022, Mannucci2022,Sandoval_2023,Wu_2024,Schwartzman2024}, which are not covered here. The completeness analysis in Section \ref{sec:fcomp} accounts for their contribution to the pair statistics.

\begin{figure*}
\centering
\includegraphics[width=\textwidth]{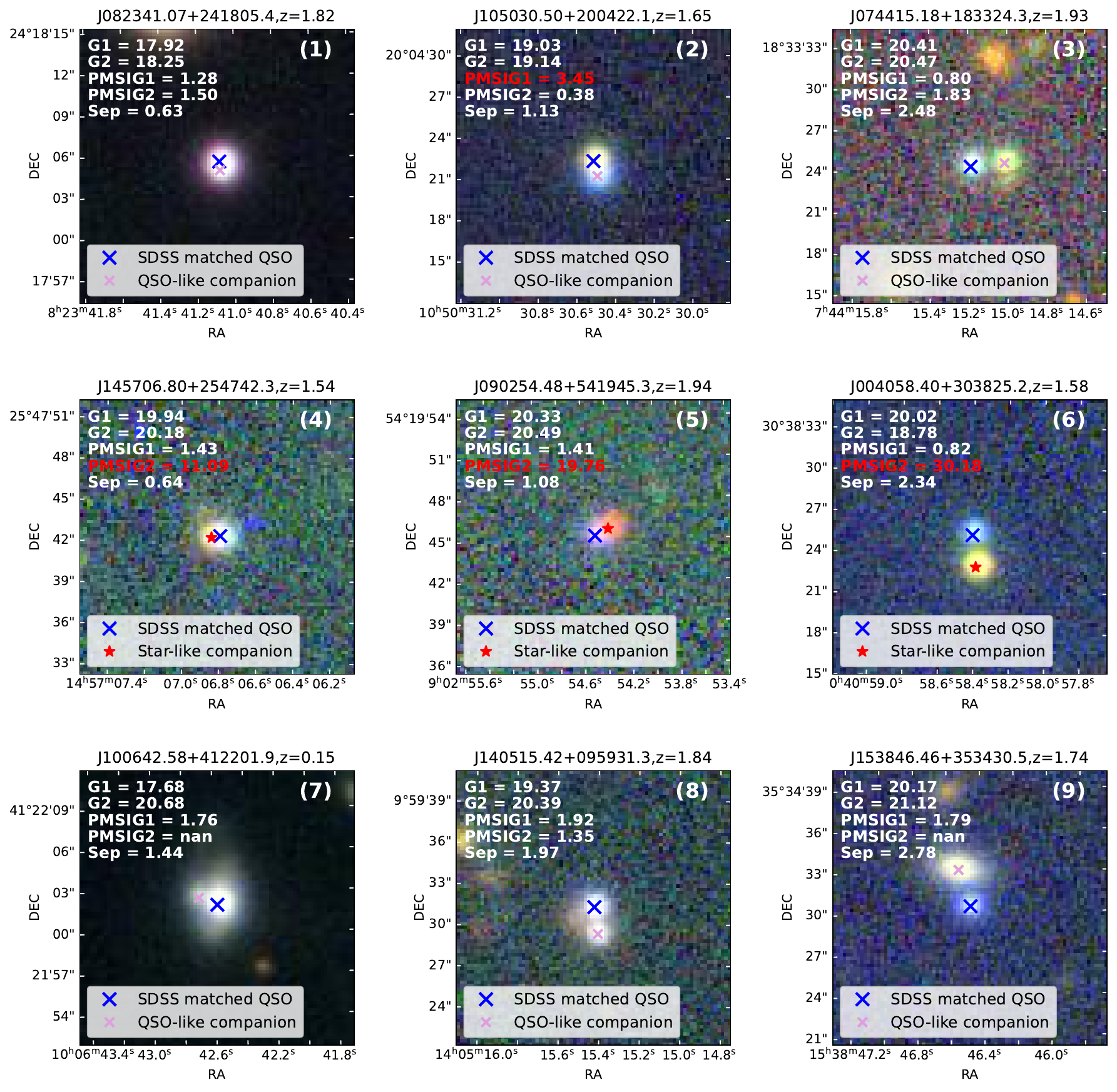}
 \caption{Pair examples of stack cutout images within $20\arcsec$ in $gri$ bands from the PanSTARRS. The SDSS name and redshift for the matched DR16Q quasar is shown on the top of each panel. The magnitude, significance of proper motion, and separation for each Gaia resolved pair matched with at least one DR16Q quasar within $3\arcsec$ are shown on the upper left of each panel. $\rm PMSIG$ larger than 3 is marked red. The blue cross, purple cross, and red star represent the matched SDSS DR16Q quasar, quasar-like companion, and the starlike companion classified by proper motion, respectively. Panels (1)-(3) display pairs at separations $<1 \arcsec$, $\sim 1 \arcsec$, and $\gg 1\arcsec$ for objects with quasar-like companions, while panels (4)-(6) show those for pairs with starlike companions. As shown in panels (1)-(6), most systems in our pair sample do have two ``nuclei" or sources if the separation is larger than typical resolution ($\sim 1 \arcsec$) for Pan-STARRS images. Panel (7) is an example for a very extended, low redshift system with quasar-like companion and no $\rm PMSIG2$ detection, indicating contamination from potential host galaxy. Panel (8) presents a lensed quasar system confirmed by follow-up observations with a lensing galaxy between two sources \citep{Lemon2019}. Panel (9) shows a quasar with a very faint companion that is classified as a quasar-like companion due to no reported proper motion. But the companion seems to be an extended source rather than a quasar, demonstrating the importance of the magnitude cut. }
 \label{fig:grid}
\end{figure*}

For each pair, the closer Gaia match is denoted as the corresponding SDSS quasar, which is generally the case. Only in a few pairs with separations less than $\sim 1 \arcsec$, the companion might dominate the SDSS optical centroid, meaning the closer Gaia match could actually be the companion. Nevertheless, this detail has minimal impact on any of our statistical analyses below. We also check cutout images from PanSTARRS \citep[e.g.,][]{Chambers2016,Flewelling2020} for pairs in our Gaia-resolved pair sample, and most of the pairs with separations $>1\arcsec$ show two resolved sources or two ``nuclei". Examples of cutout images of pairs in the $gri$ bands from PanSTARRS, with similar coverage to the Gaia $G$ band, are shown in Figure \ref{fig:grid}. These images illustrate that the two sources are always resolved when their separation exceeds the Pan-STARRS resolution limit ($\sim 1 \arcsec$). The catalog of all these pairs and their properties are compiled in Table \ref{tbl:prop}.

Not all 2,497 systems in the initial pair sample are double quasars. In fact, most of these cases are SDSS quasars with non-AGN companions, such as foreground stars \citep{Shen2023}. To eliminate these contaminants, we first apply a Gaia proper motion cut to classify the companion, which has been proven to be an effective method for distinguishing between quasars and foreground stars in previous work \citep[e.g.,][]{Lemon2019, Shen2023}. Specifically, to account for measurement uncertainties, we define the significance of proper motion ($\rm PMSIG$) following \citet{Lemon2019} as
\begin{equation}
    \rm PMSIG = \sqrt{\left(\frac{pmra}{pmra\_error}\right)^2 + \left(\frac{pmdec}{pmdec\_error}\right)^2}, 
\end{equation}
where $\rm pmra$ and $\rm pmdec$ are the proper motion in right ascension and declination direction, while $\rm pmra\_error$ and $\rm pmdec\_error$ are the corresponding standard errors. We classify the companion as a ``starlike'' companion if its proper motion is detected by Gaia with $>3\sigma$ significance; otherwise, it is classified as a ``quasar-like'' companion. {Specifically, some matched sources have no reported proper motion measurements in the Gaia DR3 catalog. Such companions are also labeled as ``quasar-like" companions. For convenience, we denote cases where proper motion measurements are unavailable in the Gaia DR3 catalog as $\rm PM = NA$ hereafter.}

Based on this proper motion classification, the parent Gaia-resolved pair sample includes 861 pairs with quasar-like companions, and 1,636 pairs with starlike companions. The classification results based on proper motion or/and other criteria mentioned below are presented in Table \ref{tbl:sample}.

\begin{figure}
\centering
\includegraphics[width=0.5\textwidth]{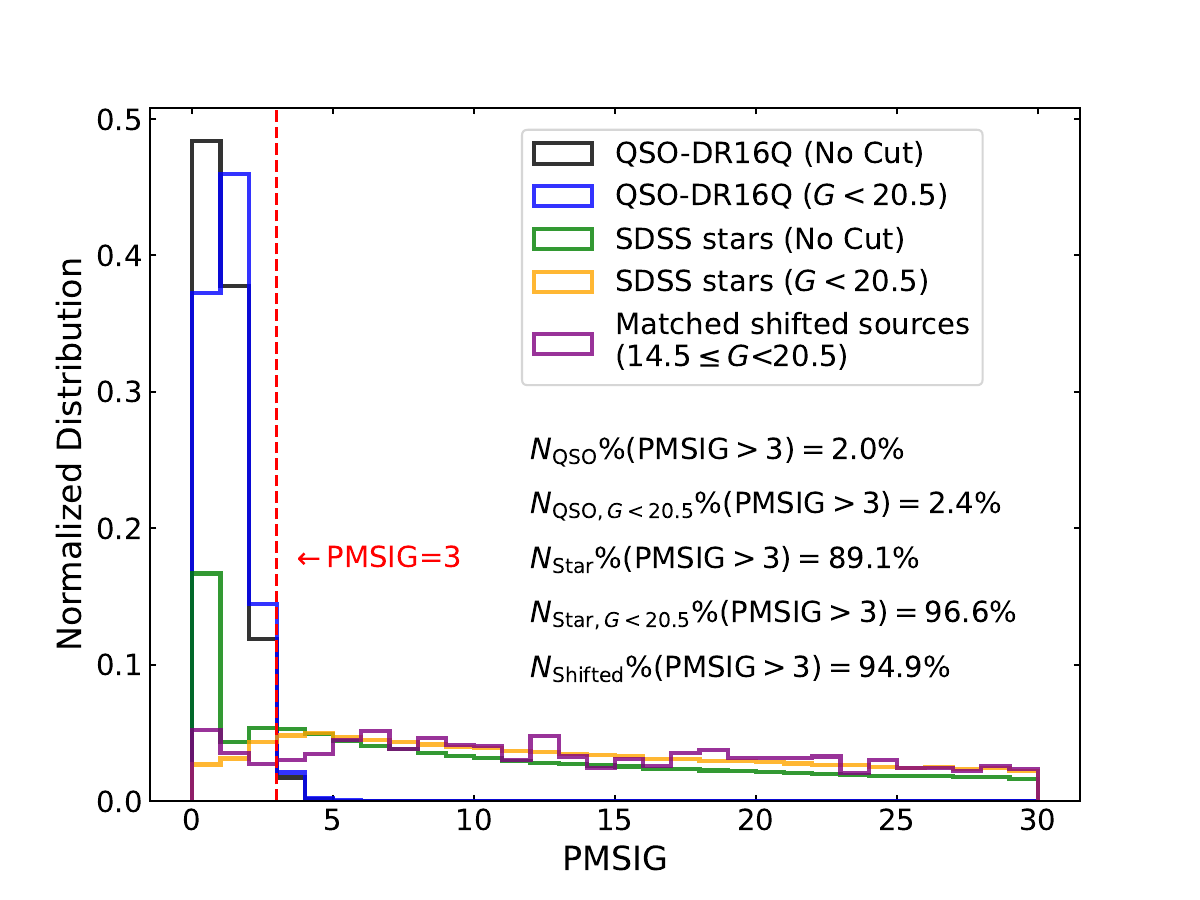}
 \caption{Distribution of the significance of proper motion (PMSIG) for different samples. The black and blue distributions represent matched SDSS DR16Q quasars without any cut and with a magnitude cut of $G<20.5$, respectively. The green and orange distributions correspond to matched SDSS stars without any cut and with $G<20.5$, respectively. The purple distribution represents newly matched sources obtained by shifting the positions of the initial matched DR16Q quasar sample by $10\arcsec$ and crossmatching with the Gaia DR3 database, as discussed in Section \ref{sec:bias}. The red dashed line represents $\rm PMSIG = 3$. Sources with no reported proper motion measurements (i.e., $\rm PM = NA$) have their $\rm PMSIG$ set to 0, and sources with $\rm PMSIG>30$ are not shown here for clarity. The fraction of sources with $\rm PMSIG>3$ for each sample is also presented. The magnitude cut here is applied to mitigate the influence of photometric incompleteness (see Section \ref{sec:match}).}
\label{fig:pmdist}
\end{figure}

Only $\sim 2\%$ of Gaia singly matched SDSS quasars have $>3\sigma$ proper motion detection, as shown in Figure \ref{fig:pmdist}. We also crossmatched SDSS spectroscopically confirmed stars with Gaia DR3, finding a very different proper motion distribution compared with that of singly matched quasars. In detail, $\sim 90\%$ of Gaia matched stars with no magnitude cut have $>3\sigma$ proper motion detection. This test indicates that our proper motion cut excludes only a negligible fraction of bona fide double quasars. However, the number of starlike companions significantly exceeds that of quasar-like companions, and some stars, especially faint stars, have unreliable Gaia proper motion measurements, and thus contaminate the double quasar sample, which is further discussed in following paragraphs and also in Sections \ref{sec:sample} and \ref{sec:bias}.

Another complication is that faint and low-redshift systems may be misclassified due to Gaia's detection limit and contamination from the extended host galaxy emission. To balance sample size and purity, we limit the redshift and magnitude of systems in the pair sample to ensure reliable identification of the double quasar population. We require each matched Gaia source to have a magnitude cut of $G<20.5$ and a redshift cut of $z>0.5$. These limits are more lenient than $G<20.25$ and $z>1.5$ adopted in \citet{Shen2023}, providing a larger sample over a broader redshift range than our earlier work. 

The criterion $G<20.5$ ensures a high completeness in Gaia detection and astrometric (e.g., proper motion) measurements. {The fractions of quasars with reported proper motion among all matched DR16 quasars in the magnitude bins $G < 20.25$, $20.25 \leq G < 20.5$, and $G \geq 20.5$ are $97\%$, $92\%$, and $41\%$, respectively.} For the companions in the pair sample only, the proper motion completeness is approximately $100\%$, $99\%$, and $50\%$ for the same magnitude ranges. Panel (9) in Figure \ref{fig:grid} presents an example of a faint companion with no reported proper motion (i.e., $\rm PM = NA$), appearing as an extended source. Furthermore, we show the proper motion distribution for matched quasars and stars with $G<20.5$ in Figure \ref{fig:pmdist} and set ${\rm PMSIG} = 0$ if no proper motion measurement is reported. The distributions for both matched SDSS quasars and stars with the magnitude cut differs significantly from that without the cut around ${\rm PMSIG} = 0$, indicating higher completeness at $G<20.5$. Additionally, after applying the magnitude cut, about 97\% of Gaia-matched SDSS stars have ${\rm PMSIG} > 3$, resulting in a purer sample for our proper-motion-based double quasar classification.

\begin{figure}
\centering
\includegraphics[width=0.47\textwidth]{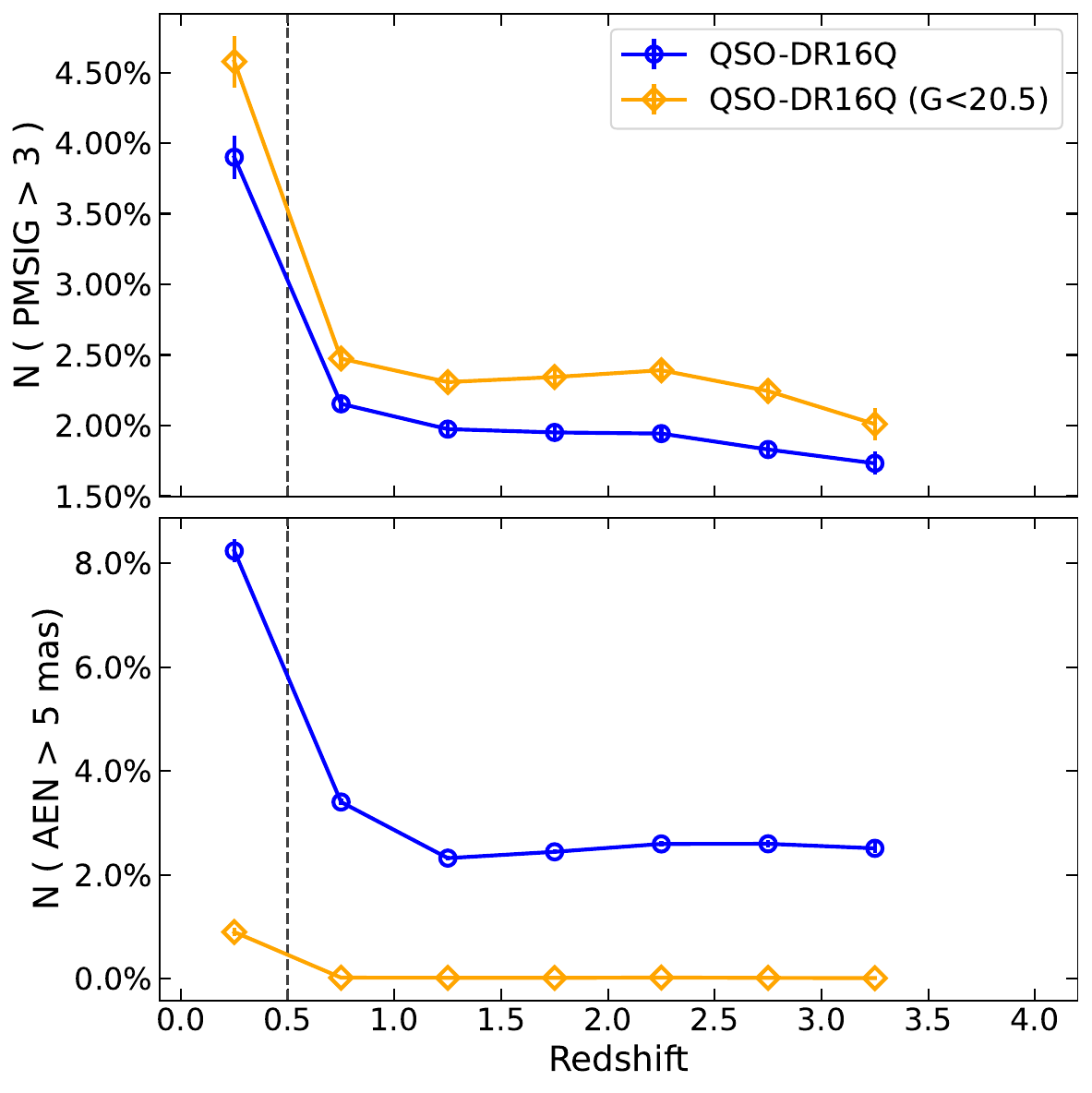}
 \caption{
 The statistical astrometric properties of DR16Q quasars matched with the Gaia DR3 database. The blue lines represent statistics for objects without a magnitude cut, while the orange lines correspond to objects with $G < 20.5$, where Gaia's detection completeness is high. The Poisson errors are displayed for each line. Top: The significance of proper motion ($\rm PMSIG$) for matched DR16Q quasars as a function of redshift. The fraction of matched quasars with high proper motion significance ($\rm PMSIG > 3$) rises rapidly at redshifts below 0.5, suggesting increased stellar contamination. Bottom: The astrometric excess noise ($\rm AEN$) for matched DR16Q quasars as a function of redshift. Similarly, the fraction of matched quasars with high astrometric excess noise ($\rm AEN > 5$ mas) increases markedly at redshifts below 0.5, indicating potential contributions from host galaxy emission. 
 }
\label{fig:dr16q}
\end{figure}

The $z>0.5$ redshift cut mitigates cases with severe contamination from the host galaxy and foreground stars. Galaxies at lower redshifts will have relatively more flux covered by the Gaia $G$ band to complicate astrometric measurements. In Figure \ref{fig:dr16q}, the upper panel shows that the fraction of singly-matched DR16Q quasars with a 3$\sigma$ proper motion detection increases rapidly at $z<0.5$, likely due to poor proper motion measurements caused by host contamination. Furthermore, the lower panel shows that the fraction of matched DR16Q quasars with astrometric excess noise ($\rm AEN$) greater than 5 mas also increases significantly at $z<0.5$, further suggesting compromised astrometric measurements due to stronger host galaxy emission within the Gaia bandpass at low redshift. These two statistics demonstrate the complexity of source detection and astrometric measurements at lower redshifts due to host galaxy emission \citep[e.g.,][]{Lemon2019, Hwang2020}. Panel (7) in Figure \ref{fig:grid} shows a low-redshift pair example with an extended structure, which may impact Gaia proper motion measurements. The imposed $z>0.5$ redshift cut ensures we have a cleaner and more complete double quasar sample than the one without a redshift restriction. 


Our $G<20.5$ flux limit roughly corresponds to a quasar bolometric luminosity $L_{\rm bol}>10^{44.5}\,{\rm erg\,s^{-1}}$ at $z>0.5$, or SDSS $i<20.38$ if we adopt a magnitude conversion of $G=i+0.12$ assuming a fixed quasar power-law continuum $f_{\nu}\propto\nu^{-0.5}$ \citep{Shen2011}. The initial DR16Q quasar sample satisfying these redshift and magnitude cuts and having single Gaia matches contains 302,940 DR16Q quasars. And the ``parent pair sample" limiting to $z>0.5$ and $G<20.5$ contains 1,120 unique pairs, including 162 quasar-like and 958 starlike companions, as listed in Table \ref{tbl:sample}. The following statistical analyses will primarily focus on the parent pair sample with $z>0.5$ and $G<20.5$, as it is considered more reliable based on the discussions above. For completeness, we also include the number of candidate double quasars and quasar-star pairs at $z\leq0.5$, or $G\geq20.5$ in Table \ref{tbl:sample}, though the purity of these candidate pairs at fainter luminosities and/or lower redshifts is significantly lower.

As a check, when we apply the same redshift and magnitude cuts, we reproduce the pair sample in \citet{Shen2023} based on Gaia EDR3. This is expected as there is minimal difference in the astrometric measurements between Gaia DR3 and EDR3.

\subsection{The Pair Sample}
\label{sec:sample}

\subsubsection{Color, Separation and Magnitude Distributions}
\label{sec:dist}

\begin{figure*}
\centering
\includegraphics[width=\textwidth]{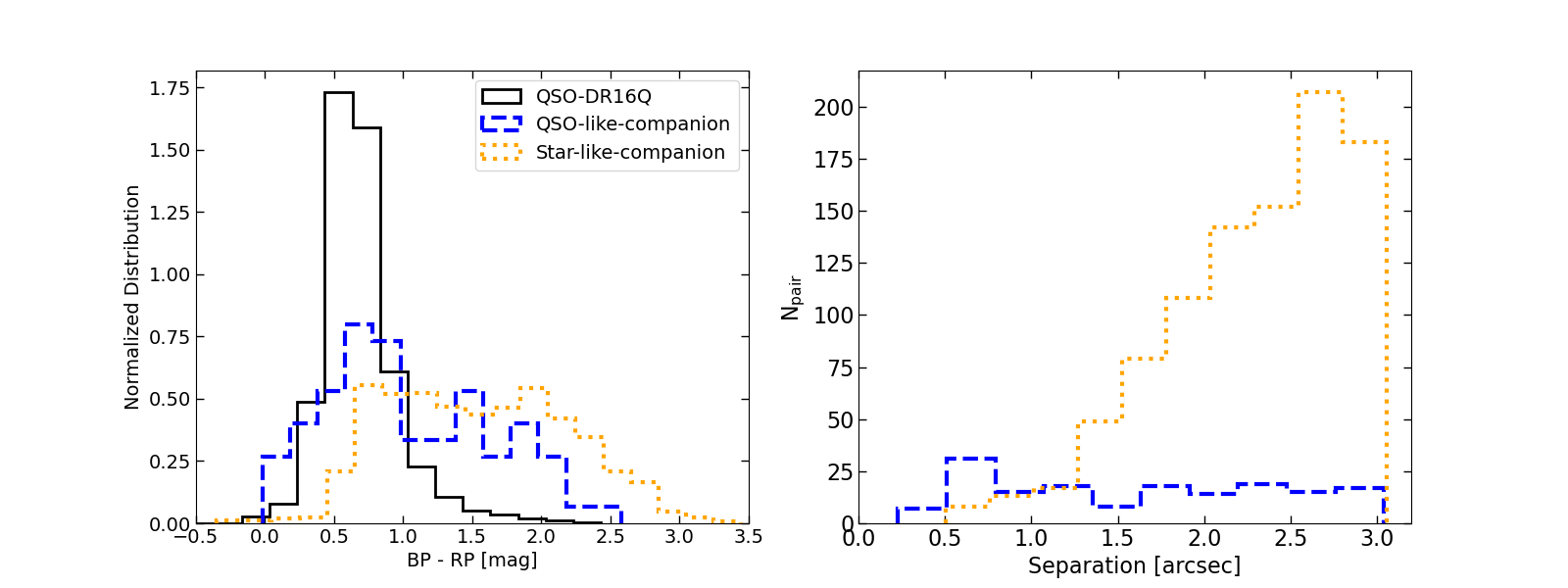}
\caption{
Distributions of the final pair sample (limited to $z > 0.5$, and $G<20.5$) in Gaia BP - RP color (left) and separation (right). These distributions are based on the raw pair statistics, without corrections for the star–quasar superpositions in $<1\farcs 5$ pairs based on the PCA and literature search, and pair incompleteness (see Section \ref{sec:pca} and Section \ref{sec:fcomp}, respectively). The color distributions (left) of matched DR16Q quasars in the pair, i.e., ``QSO-DR16Q" (black) and companions classified as ``QSO-like" based on proper-motion cut (blue) are different from that of companions classified as ``starlike" (orange), although the color distribution of ``QSO-like" companions has more fraction of red objects compared to DR16Q quasars, suggesting the existence of potential stellar contamination. The pair separation distributions (right) are also different for pairs with QSO-like and starlike companions: the pair separation distribution for star-like companions decreases rapidly at smaller separations, which is expected due to the reduced geometric cross-section for foreground superpositions; in contrast, the pair separation distribution for QSO-like companions remains relatively uniform across these separations
}
\label{fig:color_sep}
\end{figure*}

\begin{figure*}
\centering
\includegraphics[width=\textwidth]{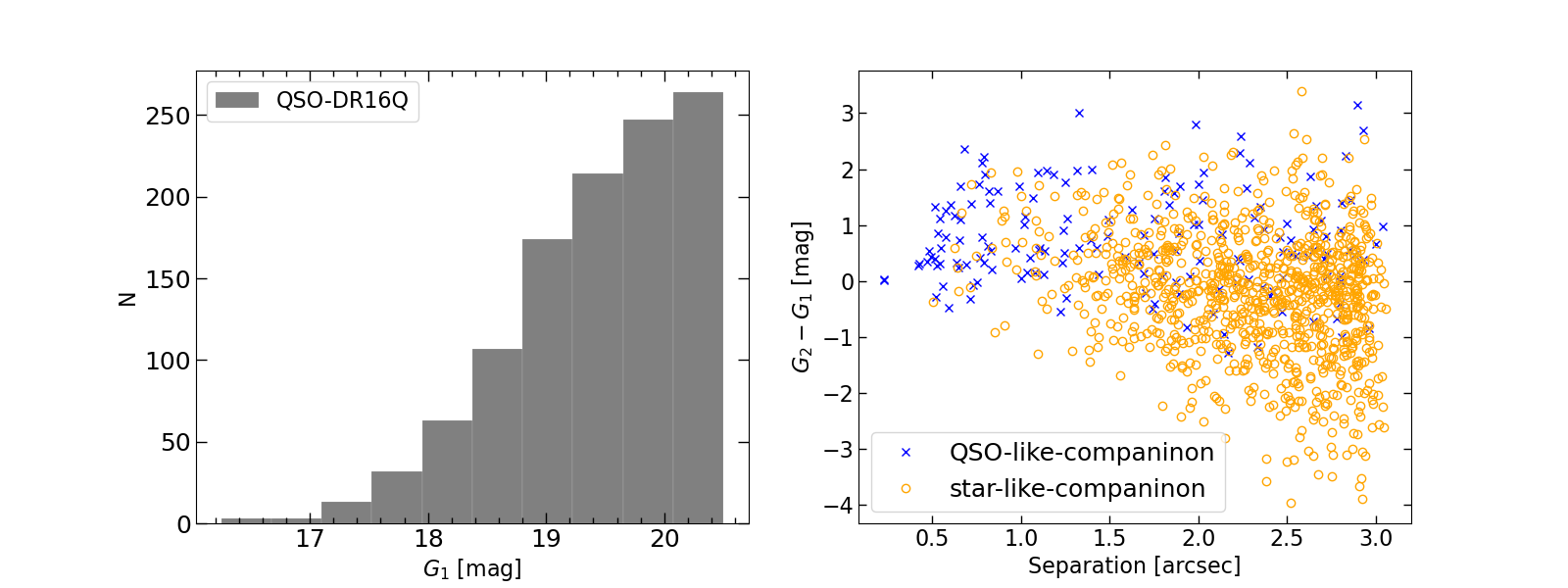}
 \caption{Statistical properties of the parent pair sample. Left: $G$-band magnitude distribution for the quasar component in DR16Q ($G1$) in each pair. Right: magnitude contrast ($G2$-$G1$) between the companion and the DR16Q quasar as a function of pair separation. At large separations, the companion can be significantly brighter than the DR16Q quasar, particularly in the case of starlike companions. Nevertheless, the majority of pairs exhibit a flux contrast of less than a factor of 10. Similar to Figure 1, the pairs shown here have not been corrected for incompleteness in the subarcsecond regime nor filtered using the PCA.}
\label{fig:contrast_sep}
\end{figure*}

To evaluate the quality of proper motion classification, several statistical properties for the parent pair sample with $z>0.5$ and $G<20.5$ are presented in Figures \ref{fig:color_sep} and \ref{fig:contrast_sep}. Figure \ref{fig:color_sep} (left) shows the distributions of Gaia BP - RP colors for the matched DR16Q quasars in pairs and their starlike and quasar-like companions from the full Gaia pair catalog. We excluded pairs with separations less than $1\arcsec$ to avoid crosstalk in color measurements, because Gaia photometry, measured within a $3\farcs5 \times 2\farcs1$ window \citep[][]{Riello2021}, could significantly affect deblending for the closest pairs.

The left panel of Figure \ref{fig:color_sep} shows that the color distribution of starlike companions, as classified by Gaia proper-motion detection, differs from that of the primary DR16Q matched quasars or quasar-like companions. Additionally, the color distributions of both DR16Q quasars in pairs and quasar-like companions are concentrated toward the blue end, although the distribution for quasar-like companions is broader. The right panel of Figure \ref{fig:color_sep} presents the distribution of pair separations for starlike and quasar-like companions. The number of starlike companions decreases rapidly at smaller separations, as expected due to the reduction in geometric cross-section and the constant sky density of the foreground (star) population. However, this trend may also be influenced by pair-resolving incompleteness at separations $\lesssim 1\arcsec$. In contrast, the distribution for quasar-like companions is relatively flat, suggesting an intrinsic population associated with the primary quasar.

Given the differences in color and separation distributions \citep{Shen2023}, the classification based on proper motion effectively removes most stellar contamination, resulting in a larger and relatively pure double quasar sample even with relaxed redshift and magnitude thresholds than those used in \citet{Shen2023}. However, the broader color distribution of quasar-like companions suggests residual stellar contamination due to the relaxed magnitude and redshift cuts, {though it may also result from a population of dust-reddened quasars in pairs with redder spectra than typical SDSS broad-line quasars.}

Figure \ref{fig:contrast_sep} displays additional statistical properties of the parent pair sample with $z>0.5$ and $G<20.5$. 
The left panel shows the magnitude distribution of the primary DR16Q mathched quasars in the pair for reference, while the right panel shows the magnitude contrast ($G_2 - G_1$) between the primary quasar and its companion at different separations. Starlike companions can be significantly brighter at large separations, which contrasts markedly with quasar-like companions. The differences in the distributions of color, separation, and magnitude contrast between quasar-like and star-like companions all indicate that the proper motion classification is relatively effective in distinguishing double quasars from quasar+star pairs.




\begin{table*}[]
\caption{Crossmatch and Classification Results\label{tbl:sample}}
\centering
\begin{tabular}{c|c|c}
\hline
\hline
{} & $z>0.5$ and $G<20.5$ & Other$^{1}$ \\ \hline
{Initial matched DR16Q QSO sample}&\multicolumn{2}{c}{492,724 DR16Q quasars$^2$} \\ \hline
{Parent matched DR16Q QSO sample} & {302,940} & {Not used in} \\ 
{for statistical analyses in Section \ref{sec:res}} & {DR16Q quasars} & {statistical analyses} \\ 
\hline
\hline
{Initial Gaia-resolved pair sample}&\multicolumn{2}{c}{2,497 unique pairs$^2$} \\ \hline
{Parent pair sample} & 162 double quasars & 699 double quasars \\ 
{(proper motion classification)} & 958 quasar-star pairs & 678 quasar-star pairs \\ \hline
{Refined pair sample} & 136 double quasars& 667 double quasars \\ 
{(by PCA analyses)} & 984 quasar-star pairs & 710 quasar-star pairs \\ \hline
{Final pair sample} & 136 double quasars & 667 double quasars \\ 
{(after literature search)} & 984 quasar-star pairs & 710 quasar-star pairs \\ 
\hline
\hline
\end{tabular}
\tablecomments{$^{1}$ ``Other" refers to low-redshift or faint sources with $z \leq 0.5$ or $G \geq 20.5$, for which classification results may be unreliable due to photometric incompleteness for faint sources and host galaxy contamination at low redshifts. Nevertheless, all crossmatch and classification results, except for matched DR16Q quasars, are included here for completeness. The faint or low-redshift DR16Q quasars are excluded from the statistical analyses in Section \ref{sec:res} and are therefore not presented here.
\\
$^2$ Numbers and descriptions spanning two columns represent sources that have not been filtered based on redshift or magnitude.}
\label{tab:example}
\end{table*}

\subsubsection{PCA Analysis and Literature Search}
\label{sec:pca}

The residual stellar contamination in our pair classification may be particularly bad at separations $\lesssim 1\arcsec$, where Gaia may struggle to resolve the proper motion measurements of such close neighbors, as the overlapping sources may significantly affect proper motion measurements by Gaia. To estimate the residual stellar contamination in quasar-like companions, we follow \citet{Shen2023} and apply a spectral principal component analysis (PCA) technique to decompose the SDSS spectrum. For completeness, this method is applied to 307 pairs with separations of $<1\farcs 5$ from the initial catalog of all 2,497 pairs. Pairs within $1\farcs 5$ separations are close enough for the SDSS fiber spectroscopy, with a diameter of $2\arcsec$ or $3\arcsec$, to capture most of the light from both components. Thus, PCA may decompose potential quasar+star superpositions using quasar and stellar PCA templates available from the SDSS website. After running PCA on these pairs, we visually inspect all PCA-decomposed spectra and flag obvious quasar+star superpositions. This step is necessary because automatic classifications from PCA decomposition are often unreliable due to degeneracies in the decomposition and noise in the spectra of these close pairs. The PCA classification results are presented in the ``PCA\_TYPE" column of Table \ref{tbl:prop}.

For the parent pair sample with $z > 0.5$ and $G < 20.5$, 155 out of 1,120 pairs have separations of $<1\farcs5$. Among these 155 pairs, 77 pairs have quasar-like companions based on the Gaia PM criterion. And among these 77 pairs, the PCA results identified 26 pairs that are apparent star superpositions, implying a residual stellar contamination rate of $34\%$ (26/77) in pairs with separations of $<1\farcs5$. This contamination rate is consistent with that estimated in \citet{Shen2023} for the high-redshift subset. We also visually inspect the PCA decomposition for starlike companions (based on PMSIG) in pairs with separations of $<1\farcs5$, and most of them show composite spectra similar to stars or galaxies, further supporting the effectiveness of the Gaia PMSIG classification. To mitigate star contamination in the $<1\farcs5$ regime, we exclude these PCA-identified quasar+star pairs from the double quasar sample, resulting in 136 pairs with quasar-like companions and therefore 984 starlike companions in the ``refined pair sample", as summarized in Table \ref{tbl:sample}. However, we cannot fully eliminate stellar contamination in pairs with separations of $>1\farcs5$, and we defer a rough estimation to Section \ref{sec:bias} to assess the overall purity of the refined pair sample at these large separations. 

The PCA results for low-redshift or faint quasar or quasar-star pairs that do not meet our statistical cuts are also visually inspected. This includes 152 pairs with ($z\leq0.5$ or $G\geq20.5$) and separations of $<1\farcs5$. The PCA results show that approximately $29\%$ of quasar pairs (32/111) with separations of $<1\farcs5$ have in fact star superpositions. However, these results are not entirely reliable, as some faint sources do not have significant contribution to the total SDSS spectrum, and Gaia's PM detection accuracy is limited for faint or low-redshift close pairs. Therefore, while we present the PCA results for these low-redshift or faint targets in Tables \ref{tbl:sample} and \ref{tbl:prop}, we exclude them from further statistical analysis.

After the removal of the foreground star superpositions, most of the remaining 136 pairs with $z>0.5$ and $G<20.5$ should be double quasars, including genuine dual quasars and lensed quasars. Although many gravitationally lensed quasars have been identified through various surveys, such as Gaia and follow-up observations \citep[e.g.,][]{Lemon2019,Lemon2023}, the completeness of catalogs of confirmed lensed quasars is difficult to quantify due to the challenge of distinguishing them from physically associated dual quasars \citep[e.g.,][]{Chen2023a,Li2023,Gross2023}. Consequently, it is not possible to fully exclude lensed quasars from our sample or remove their influence on statistical analyses.

Nevertheless, we attempt to identify lensed quasars within our Gaia-resolved pairs sample. First, we compile a catalog of lensed quasars from the literature, which includes entries from the Gravitationally Lensed Quasar Database \citep{Lemon2019} as well as other newly confirmed lensed quasars reported after 2019 \citep[][]{Delchambre2019,Lemon2020, Lemon2023,Hawkins2021,Jaelani2021,Stern2021,Chan2022,Desira2022,Li2023, Napier2023,DusanTubin2023,Dux2023, Dux2024,Schwartzman2024}. It is possible that additional sources observed in the literature are missing from these resources. For completeness, we cross-match all 2,497 unique pairs from the initial pair sample with this compiled lensed quasar catalog. We identify 59 pairs among the 2,497 as lensed systems based on follow-up observations, 40 of which meet the redshift and magnitude thresholds of $z>0.5$ and $G<20.5$. These publicly reported cases are labeled in the ``KNOWN" column in Table \ref{tbl:prop}, with references to their source literature provided in the ``REFERENCE" column.

Among the 136 quasar-like companions at $z>0.5$ and $G<20.5$, classified via proper motion and PCA inspection, 36 are known lensed quasars in the compiled lens catalog. Additionally, four systems classified as quasar-star pairs based on our Gaia proper motion criterion and PCA inspection are in fact confirmed lensed quasars. Below we examine these four mis-classified systems in detail:

\begin{itemize}
    \item J1355306.34+113804.7 (with a $1\farcs39$ separation) has a companion with a proper motion significance of $\rm PMSIG2 \approx 6$, as also reported in \citet{Shen2023}. For comparison, the median proper motion significance for pairs classified as quasar-star systems is approximately 25. This may be a rare case where Gaia’s proper motion measurement is influenced by a nearby bright neighbor.
    
    \item J095122.57+263513.9 and J133222.62+034739.9 (with separations of $1\farcs10$ and $1\farcs12$, respectively) have companions with $\rm PMSIG2$ values of 3.2 and 3.4, slightly above the threshold of 3. These two systems are on the boarderline of the selection criterion and may represent quasar-like companions missed by the proper motion cut. However, only $\sim 2\%$ of Gaia singly matched quasars have $>3\sigma$ proper motion detections, as mentioned in Section\ref{sec:match}, and only $\sim 2\%$ of companions fall within the $3 \leq {\rm PMSIG2} \leq 4$ range. Therefore, this effect is unavoidable but negligible for the overall statistics. 
    
    \item The final system, J165043.44+425149.3, has a companion with $\rm PMSIG2 = 0.96$, which suggests it should be classified as a double quasar system. However, the PCA results show some plausible stellar features, leading to its initial reclassification as a quasar-star pair after inspection. This system is matched with a known lensed quasar system, and the misclassification may have been caused by contamination from the foreground lensing galaxy located between the two lensed quasars \citep{Morgan2003}. 
\end{itemize}

As our double quasar sample is primarily selected based on the PMSIG criterion, to maintain homogeneous selection, we opt to only reinstate the last system (J165043.44+425149.3) in our double quasar sample since it has $\rm PMSIG < 3$. Additionally, we excluded J003337.58+201538.1 (with a $1\farcs69$ separation), a reported {quasar-star pair} \citep{More2016}, from the double quasar sample for further analysis. As a result, the ``final pair sample" with $z>0.5$ and $G<20.5$ has 136 double quasars and 984 quasar-star pairs as reported in Table \ref{tbl:sample}. And the final classification decision for each pair is provided in the ``TYPE" column.

\begin{figure}
\centering
\includegraphics[width=0.5\textwidth]{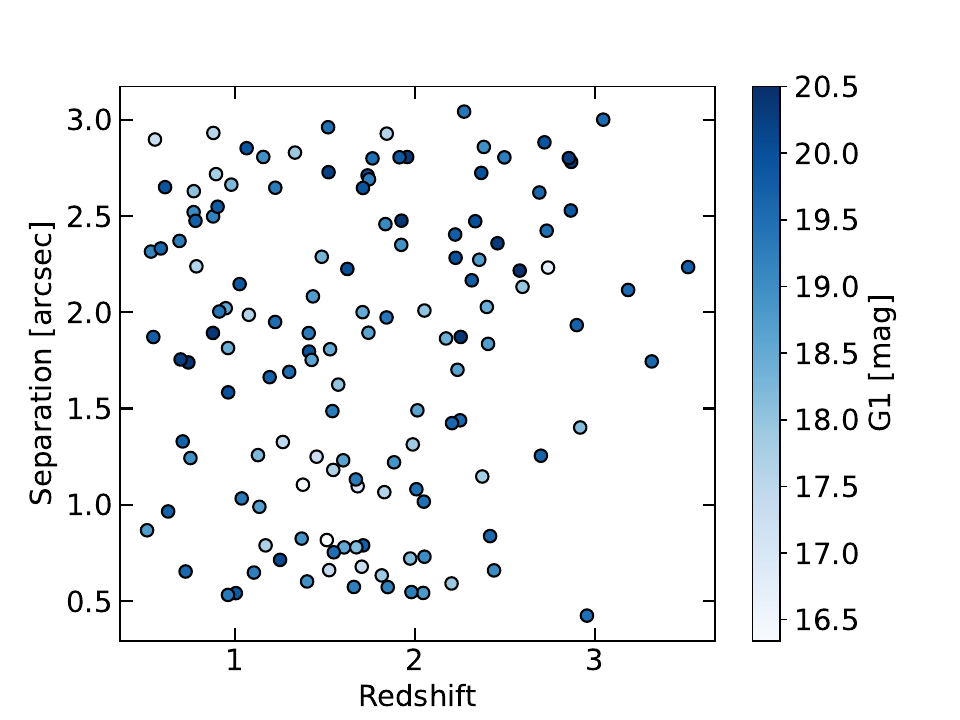}
 \caption{Distribution of our final sample of 136 double quasars in the redshift-separation space, color-coded by the $G1$ magnitude (as listed in Table 1) of each pair. }
\label{fig:zsep}
\end{figure}

Figure \ref{fig:zsep} shows the distribution of the 136 pairs of final double quasar sample in redshift-separation space. The double quasars have pair separations ranging from $0\farcs4$ to $\sim 3 \arcsec$, forming the basis of our subsequent analyses. At the sample's median redshift of $z=1.7$, these pairs correspond to projected separations of $3 \ {\rm kpc} \lesssim r_p \lesssim 30 \ {\rm kpc}$, i.e., on galactic scales. While individual pairs may have 3D separations exceeding 30 kpc, this population statistically traces the radial distribution of quasar pairs. 

\begin{deluxetable*}{cccc}
\tablecaption{Pair Sample Data}
\label{tbl:prop}
\tablehead{
\colhead{Column} & \colhead{Format} & \colhead{Units} & \colhead{Description}
}
\colnumbers
\startdata
SDSS\_NAME & STRING &  & J2000 hhmmss.ss $\pm$ ddmmss.s \\ 
Z & DOUBLE &  & Systemic redshift in \\
 & & &  \citet{Wu2022} \\
PLATE &  &  & Plate number (SDSS spec) \\
FIBERID &  &  & FiberID (SDSS spec) \\
MJD &  &  & MJD (SDSS spec) \\
GAIA\_RA1 & DOUBLE & deg & Gaia RA \\
GAIA\_DEC1 & DOUBLE & deg & Gaia DEC \\
GAIA\_RA2 & DOUBLE & deg & Gaia RA \\
GAIA\_DEC2 & DOUBLE & deg & Gaia DEC \\
G1 & DOUBLE & mag & Gaia G mag \\
G2 & DOUBLE & mag & Gaia G mag \\
BP\_RP1 & DOUBLE & mag & Gaia BP-RP color \\
BP\_RP2 & DOUBLE & mag & Gaia BP-RP color \\
PM\_SIG1 & DOUBLE &  & PM significance \\
PM\_SIG2 & DOUBLE &  & PM significance \\
PAIR\_SEP & DOUBLE & arcsec & Pair separation \\
KNOWN & STRING &  & Literature classification \\
REFERENCE & STRING &  & Related literature \\
PCA\_TYPE & STRING & & PCA classification \\
TYPE & STRING &  & Final pair classification \\
F\_COMP & DOUBLE &  & Pair Completeness (Section \ref{sec:fcomp})\\
\enddata
\tablecomments{
For each pair, the DR16Q quasar is designated as index 1 and the companion as index 2, regardless of their brightness, which means the quasar may sometimes be dimmer than its companion, particularly when the pair separation is large. Gaia measurements are from DR3 (null values are masked). The column ``REFERENCE" provides the source papers for the identified pairs, including those selected through previous studies and confirmed via follow-up observations. The matched lensed quasars from the Gravitationally Lensed Quasar Database \citep{Lemon2019} are all labeled as ``Lemon2019", even when the lensed quasars were originally collected from other literature sources rather than directly from \citet{Lemon2019}. The column ``PCA\_TYPE" refers to the PCA classification results for pairs at separations of $<1\farcs5$ (the PCA results for pairs at larger separations are masked), while the column ``TYPE" indicates the final pair classification. ``QQ" refers to a double quasar, ``$\text{QS\_PM}$" refers to a quasar+star pair based on proper motion, and ``$\text{QS\_PCA}$" refers to a quasar+star pair based on spectral PCA; one quasar (J0033+2015) is a known quasar+star pair \citep{More2016}, and we set its TYPE = ``$\text{QS\_KNOWN}$". The associated FITS file is available in the online version of this paper.
}
\end{deluxetable*}

\subsection{Foreground Star Superposition Rate}
\label{sec:bias}

\begin{figure}
\hspace{-20pt}
\includegraphics[width=0.5\textwidth]{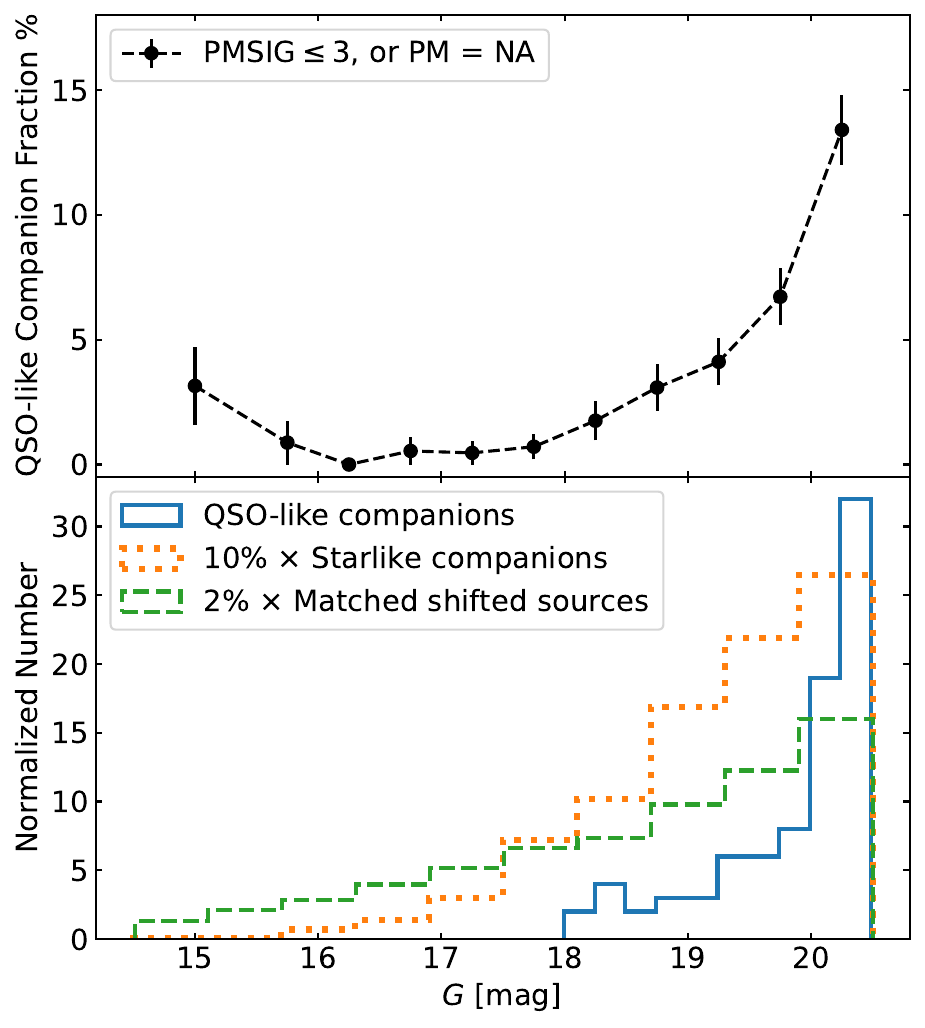}
 \caption{Upper panel: Fraction of newly matched sources with insignificant proper motion (i.e., $\rm PMSIG\leq3$, or $\rm PM = NA$) as a function of $G$-band magnitude. These sources are obtained by randomly shifting the positions of the initial matched DR16Q QSO sample by $10\arcsec$ and cross-matching with the Gaia DR3 database (see details in \ref{sec:bias}). A magnitude cut of $14.5<G<20.5$ is applied, and the Poisson error for each magnitude bin is presented. Lower panel: The $G$-band magnitude distributions of different samples. The blue solid line represents the distribution of QSO-like companions ($\rm PMSIG \leq 3$, or $\rm PM = NA$) in the refined pair sample, selected with $z>0.5$, $G<20.5$, and $\rm PAIR \ SEP > 1\farcs 5$. The orange dotted line shows the normalized distribution of starlike companions ($\rm PMSIG > 3$) with the same selection criteria. The green dashed line corresponds to the normalized $G$-band magnitude distribution of newly matched shifted sources, where an additional magnitude cut of $14.5<G<20.5$ is applied for consistency. Normalization factors of 10\% and 2\% are applied to the distributions of starlike companions and matched shifted sources, respectively, to roughly match the total number of QSO-like companions in the refined pair sample for clarity.
 }
\label{fig:pm_and_gmag}
\end{figure}

The PCA analysis effectively removes most stellar contamination for pairs with separations less than $1\farcs5$, as they fall within a single SDSS fiber. However, for wider pairs, PCA cannot estimate contamination due to the lack of fiber coverage. While Gaia’s resolution largely ensures reliable proper motion measurements for such pairs, foreground contamination can still be significant. {Although only a small fraction of stars have insignificant proper motion measurements ($\rm PM = NA$, or $\rm PMSIG \leq 3$),} the number of foreground stars far exceeds that of intrinsic quasar companions, as shown in Table \ref{tbl:sample} and Figure \ref{fig:pm_and_gmag}, potentially leading to high contamination.

To estimate stellar contamination, we perform a random offset test by shuffling SDSS quasar positions in the input DR16Q sample by $10\arcsec$ and searching for Gaia sources within a $3\arcsec$ radius. This test preserves the foreground stellar density distribution relevant to the SDSS quasar sample. We assume the newly matched sources represent the foreground superpositions, calculate their proper motion significance, and apply a magnitude cut of $14.5 < G < 20.5$. As shown in the lower panel of Figure~\ref{fig:pm_and_gmag}, the magnitude cut mimics the magnitude distribution of companions in the parent pair sample. Moreover, the proper motion significance distribution of newly matched offset sources closely matches that of the SDSS star catalog for $G<20.5$, as shown in Figure~\ref{fig:pmdist}, suggesting that the matched sources in the shuffle test are predominantly stars. We also present the fraction of newly matched sources with insignificant proper motion as a function of magnitude in the upper panel of Figure \ref{fig:pm_and_gmag}. Both the fraction of newly matched sources with insignificant proper motion and the number of starlike companions rise at fainter magnitudes, highlighting the unavoidable foreground contamination despite the proper motion cut.

It is challenging to precisely quantify the star contamination in the wide separation pairs in our refined quasar pair sample. Here, we provide an approximate estimate of the stellar contamination for pairs with separations larger than $1\farcs5$. 

We derive the fraction of matched sources with significant proper motion in the initially matched DR16 QSO sample over the total number of sources in the sample. And we also compute this fraction for the newly matched sample constrained by $14.5 < G < 20.5$ in shuffle test to represent the significant proper motion ratio of the foreground sample. The two ratios are denoted as $a_1$ and $a_2$ respectively. Then, we count the number of double quasars and quasar-star pairs in the refined pair sample with $z>0.5$, $G<20.5$, and $\rm PAIR \ SEP \ge 1\farcs5$ as $N_{\rm PMSIG>3}$ and $N_{\rm INSIG}$. The expected numbers of double quasars and quasar-star pairs ($N_q$ and $N_s$ respectively) are
\begin{equation}
    \begin{cases}
        N_{\rm PMSIG>3} =  a_1 N_q + a_2 N_s \ , \\
        N_{\rm INSIG} = (1-a_1) N_q + (1-a_2) N_s \ ,
    \end{cases}
\end{equation}
where $a_1 = 97.7\%$, $a_2 = 94.9\%$, $N_{\rm PMSIG > 3} = 85$, and $N_{\rm INSIG} = 965$. The estimated contamination ratio for quasar-like companions with $z>0.5$, $G<20.5$ and $\rm PAIR \ SEP\ge 1 \farcs5$ in refined pair sample is $(1-a_2)N_s/N_{\rm INSIG} = 56 \%$, which is comparable to other works searching for dual quasar candidates but using different methods \citep[e.g.,][]{Silverman2020,Li2024}. Since this is only a rough estimation, we do not apply the correction to our large-separation pairs. But we note that the actual number of double quasar candidates in larger separations may be lower due to this uncorrected stellar contamination. This estimated contamination rate is higher than the $\sim 30\%$ for $<1\farcs5$ pairs because the larger separations enclose substantially more superposition stars. 

\subsection{Pair Completeness}
\label{sec:fcomp}

Gaia's ability to resolve pairs is incomplete, particularly in the subarcsecond regime, which significantly impacts direct statistical analyses as a function of separation. Gaia can resolve pairs with a resolution of approximately $0\farcs2$ in the along-scan direction, which may be partially compensated by multiple scans from different directions. However, deblending two very close sources remains challenging for Gaia due to its $3\farcs5 \times 2\farcs1$ photometry window. Therefore, a correction for pair-resolving completeness is required before performing statistical analyses based on pair separation.

\citet{Shen2023} calculated the pair completeness as a function of the magnitude of the brighter primary source ($G_{\text{pri}}$), the magnitude difference between the two sources ($\Delta G = G_{\text{pri}} - G_{\text{sec}}$, where $G_{\text{sec}}$ is the $G$-band magnitude of the secondary source), and their angular separation ($\Delta \theta$), using Gaia EDR3 data. The photometric and astrometric measurements remain nearly unchanged from EDR3 to DR3, so the pair completeness functions from \citet{Shen2023} are still applicable to our DR3 pair sample. The pair completeness $1/f_{\text{comp}}$ as a functions of $G_{\text{pri}}$, $\Delta G$, and $\Delta \theta$ is also available, as an electronic FITS table with its content described in Table 2 from \citet{Shen2023}. 


\section{Results}
\label{sec:res}
 

We have classified 136 double quasars and 984 quasar-star pairs in the final pair sample of 1,120 unique pairs with $z>0.5$ and $G<20.5$ after cross-matching the SDSS DR16Q catalog with GAIA DR3 data within $3\arcsec$, proper motion classification, PCA analysis and literature search. Here, we further correct the number of quasar pairs in our final double quasar sample with $z>0.5$ and $G<20.5$ using the completeness correction mentioned in Section \ref{sec:fcomp} and perform several analyses based on this final double quasar sample as follows.

To match with previous studies \citep[e.g.,][]{Shen2023}, we define the pair fraction as the ratio of the number of pairs to the total number of parent quasars with matched flux limit and redshift range. Because the pair fraction of luminous quasars is $\ll 1\%$, we neglect the small contribution of companion quasars in the parent quasar count. 

\subsection{Double Quasar Fraction}

We first calculate the completeness-corrected double quasar fraction. The completeness-corrected double quasar fraction is defined as the ratio of the number of quasar pairs in each separation bin to the total number of quasars in the parent sample, where each observed quasar pair is weighted by $1/f_{\text{comp}}$ according to its $G_{\text{pri}}$, $\Delta G$, and $\Delta \theta$. The correction is significant primarily in the subarcsecond regime \citep{Shen2023}. Here, we define the parent quasar sample as all matched SDSS DR16Q quasars in Gaia with $z>0.5$ and $G<20.5$ (302,940 quasars), the same magnitude and redshift cuts as our final pair sample. {Double quasars with separations of $0-3\arcsec$ is negligible compared to the matched SDSS DR16Q quasar population ($10^{-3} \sim 10^{-4}$ calculated in the following paragraphs), so we do not add the potential quasar-like companions to the parent sample. Moreover, the parent sample only serves as the denominator in the pair fraction calculation and does not affect the relative fraction and overall evolution trend in different separation bins.} 

Only two pairs in our double quasar sample slightly exceed the 3D grid of $G_{\text{pri}}$, $\Delta G$, and $\Delta \theta$ used to calculate $1/f_{\text{comp}}$, as their companions are too dim and their magnitude differences slightly exceed 3. J083056.16+062412.8 has a companion with $G_{\text{pri}} = 17.48$, $\Delta G = 3.01$, and $\Delta \theta = 1\farcs33$, while J113625.42+100523.2 has a companion with $G_{\text{pri}} = 17.30$, $\Delta G = 3.16$, and $\Delta \theta = 2 \farcs90$. These excesses are not significant, and their separations are beyond the subarcsecond regime; thus, we use the $1/f_{\text{comp}}$ in bins of $2.5 < \Delta G < 3.0$ as a reasonable estimate for these two pairs. 

We further estimate the uncertainties of the corrected pair statistics using bootstrap resampling of the pairs, which align well with Poisson uncertainties estimated from the raw pair counts in each separation bin. The corrected pair statistics is reported in Table \ref{tbl:stats}, and Figure \ref{fig:pair_frac} displays the completeness-corrected double quasar fraction as a function of angular separation in different redshift ranges. We adopt bootstrap errors in this analysis as they better reflect the sample characteristics, while Poisson errors for each separation bin are also provided in Table \ref{tbl:stats}. The cumulative pair fraction within $0\farcs3-3\farcs1$ is $5.7^{+0.3}_{-0.3} \times 10^{-4}$ among $z>0.5$ quasars. We further divide these quasar pairs to two groups at $z=1.5$, which is close to the median redshift of the pair sample ($\langle z \rangle=1.7$). The low-redshift group ($\langle z \rangle=1.0$) includes 54 quasar pairs with a cumulative pair fraction of $4.9^{+0.2}_{-0.2} \times 10^{-4}$, while the high-redshift group ($\langle z \rangle=2.0$) has 82 quasar pairs with a cumulative pair fraction of $6.2_{-0.5}^{+0.5} \times 10^{-4}$. The cumulative pair fraction for the higher-redshift group is only marginally larger that that for the lower-redshift group, with a difference of $0.10_{-0.03}^{+0.03}$ dex. Furthermore, Figure \ref{fig:pair_frac} shows that the double quasar fractions are consistent in most separation bins for different redshift ranges, overlapping within $1\sigma$. The only difference occurs at $\Delta \theta = 0\farcs3 - 0\farcs5$ due to no double quasar detection for the low-redshift group. 


In Figure \ref{fig:pair_frac}, the completeness-corrected double quasar fraction increases toward smaller separations, consistent with previous results using Gaia EDR3 for $z>1.5$ and $G<20.25$ quasars \citep{Shen2023}. We note that the actual fractions at separation larger than $1 \farcs 5$ should be even lower due to uncorrected stellar contamination as discussed in Section \ref{sec:bias}. And if we assume a stellar contamination rate of 56\% for the corrected number of quasar pairs at $z > 1.5$, $G < 20.5$, and $\rm PAIR \ SEP \ge 1\farcs5$ as estimated in Section \ref{sec:bias}, the cumulative double quasar fractions may decrease by $\sim30\%$, though the overall increasing trend toward smaller separations remains unchanged. The steepening of the pair fraction at small separations suggests an increase in small-scale quasar clustering on physical scales of $\lesssim 30$ kpc compared with the power-law extrapolation of the large-scale clustering \citep[e.g.,][]{Shen2023}.

\subsection{Comparison with Previous Work}

Due to the larger redshift coverage from $z \sim 0.5$ to $z \sim 4.5$, we also investigate the redshift evolution of the dual quasar fraction to compare with other observations and simulations. The definition of the dual quasar fraction as a function of redshift is similar to the double quasar fraction as a function of separation, but evaluated for physically-associated quasar pairs (as defined in Section \ref{sec:intro}) in different redshift bins for all pairs at $<3\arcsec$ separations. 

\begin{figure}[!t]
    \centering
    \includegraphics[width=\linewidth]{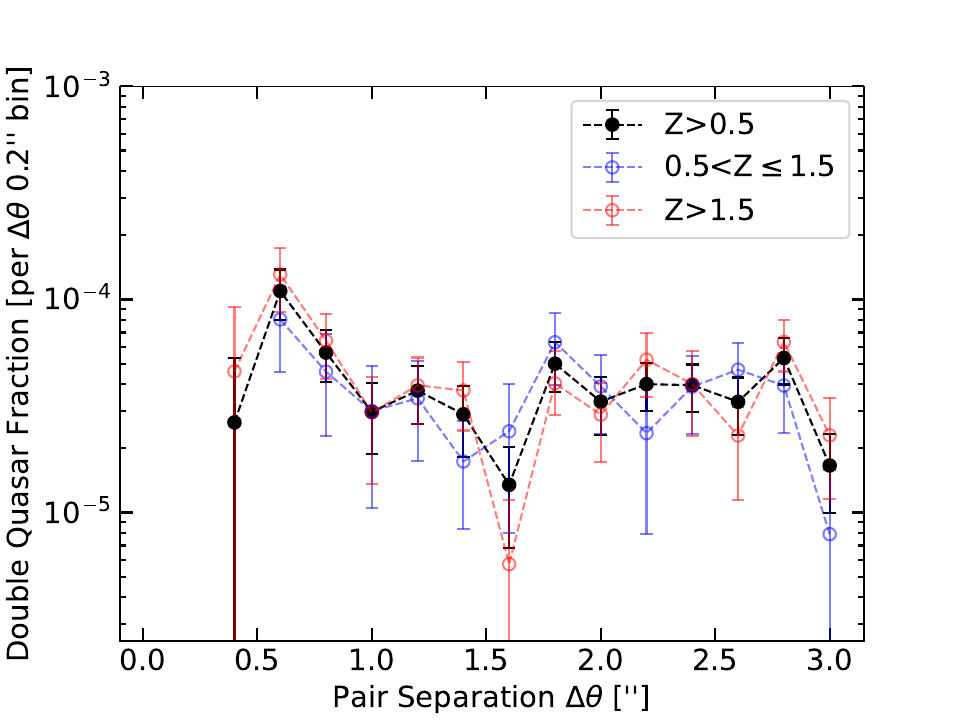}
    \caption{Measured double quasar fraction corrected for pair incompleteness (Section \ref{sec:fcomp}) at different redshift bins and $G < 20.5$. The estimated lensed quasar fraction from \citet{Shen2023} is not applied here to exclude potential lensed quasars. The black circles shows the fraction at $z>0.5$. And the blue and red circles present the fraction at lower redshift bin $0.5<z\leq1.5$ and higher redshift bin $z>1.5$, respectively. The bootstrap errors are provided for the sample in each redshift range. No double quasars are found in the $0 \farcs 3 <\Delta \theta < 0 \farcs 5$ bin for the lower-redshift group, so the fraction is zero and not shown. 
    }
    \label{fig:pair_frac}
\end{figure}

\begin{figure}[!t]
    \centering
    \includegraphics[width=\linewidth]{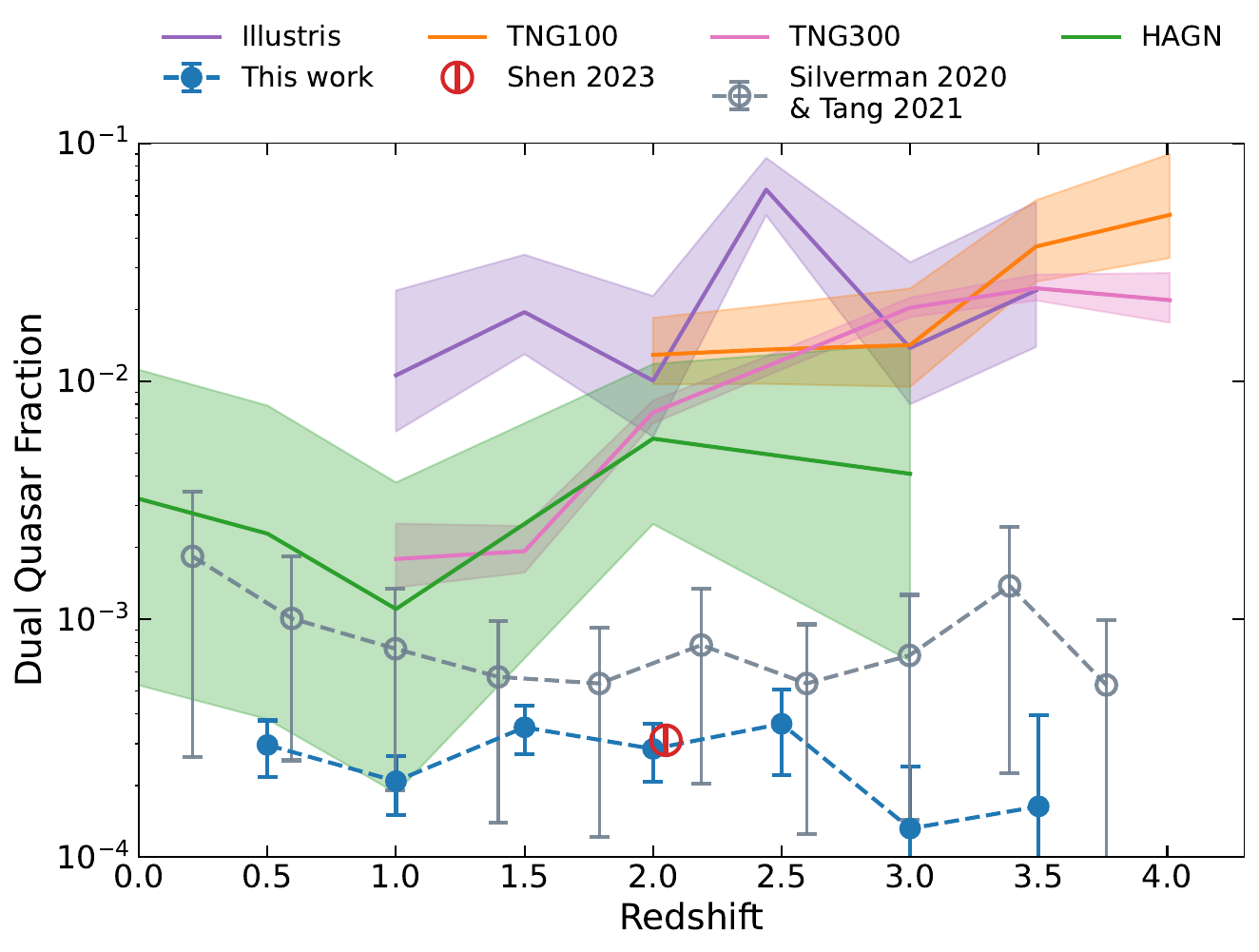}
    \caption{
    Dual quasar fraction as a function of redshift. 
    The blue data points represent the double quasar fraction from this work, compared with previous observational studies, including \citet{Silverman2020} (gray) and \citet{Shen2023} (red). 
    Dual quasar fractions in this work and \citep{Shen2023} are corrected for the lensed quasar fraction, while that in \citet{Silverman2020} are account for the spectroscopic confirmation rate and incompleteness at small separations \citep{Silverman2020,Tang2021}.
    For comparison, fractions from 4 simulations constrained by the same observational criteria as \citet{Silverman2020} are also displayed here, including Illustris (purple), TNG100 (orange), TNG300 (pink), and HAGN (green), with fractions derived by \citet{Clara2025}. The error bars or shaded regions indicate the Poisson uncertainties for each dataset.
    }
    \label{fig:zevolution}
\end{figure}

To make a fair comparison with other studies, we attempt to statistically correct for the lensed quasar contamination in the final double quasar sample. An accurate estimation of the lensed quasar contribution in the pair sample is currently infeasible, as it requires detailed follow-up observations. Here we simply assume a dual-to-lens ratio of 1:1, as suggested by some models \citep{Oguri2010}. This ratio is a statistical correction for comparison purposes and may differ from the actual lens fractions, leading to some uncertainties. However, if the true dual:lens ratio were 2:1 or 1:2, the corrected dual fractions would shift up or down only by $33\%$, and our conclusions on the redshift trend remain the same. The corrected dual quasar fractions for pair completeness and lensed quasar contamination as a function of redshift are shown as the blue curve in Figure \ref{fig:zevolution}. To ensure consistency with previous studies (discussed below), all uncertainties in Figure \ref{fig:zevolution} are represented as Poisson errors rather than bootstrap uncertainties. We set the bin size to $\Delta z = 0.5$ for double quasars with $z \leq 3.5$ and combine those at $z > 3.5$ into a single bin due to their small numbers. Figure \ref{fig:zevolution} shows weak redshift evolution from $z = 0.5$ to $z = 3.5$, consistent with our previous findings in Figure \ref{fig:pair_frac}. 

We also compare our results with other observational studies that also target luminous unobscured quasar pairs \citep{ Silverman2020,Shen2023}. \citet{Shen2023} selected double quasars using a similar method to our study but used Gaia EDR3 data with different redshift and magnitude thresholds ($z > 1.5$ and $G < 20.25$), {and found a cumulative double quasar fraction (uncorrected for the dual-to-lens ratio) of $6.2 \pm 0.5 \times 10^{-4}$ among $z>1.5$ quasars, as shown in Figure \ref{fig:zevolution} but corrected for a dual-to-lens ratio of 1:1 to estimate the dual quasar (physically-associated quasar pairs) fraction. \citet{Shen2023} also divided their sample at $\langle z \rangle = 2$ and presented double quasar fractions of $6.6 \pm 1.2 \times 10^{-4}$ and $5.9\pm1.0 \times 10^{-4}$ in the lower ($\langle z \rangle = 1.7$) and higher ($\langle z \rangle = 2.4$) redshift bins, which also shows no strong redshift evolution. The results in \citet{Shen2023} are similar to this work in both fraction and redshift evolution trend, suggesting minimal redshift evolution}

\citet{Silverman2020} identified dual quasar candidates around luminous SDSS quasars using Subaru HSC imaging, employing color selection. Briefly speaking, luminous dual AGNs at $z \leq 3.5$ with separations of 5--30 kpc are selected, where the primary AGN has $L_{\rm bol} \geq 10^{45.3}$ erg/s, and the secondary AGN (candidate) contributes at least $10\%$ of the primary's luminosity ($L_{bol} > 10^{44.3}$ erg/s). \citet{Silverman2020} reported a success rate of 3 dual quasars out of 6 candidates, but a later follow-up \citep{Tang2021} identified 3 dual quasars out of 26 additional sources. The plotted double quasar fraction adopted from their work is corrected by the the overall success rate (6 dual quasars out of 32 candidates) based on the follow-up study in \citet{Silverman2020} and \citet{Tang2021}. 


The results from \citet{Silverman2020} and \citet{Tang2021} are generally consistent with our findings within $1\sigma$, and also indicate little redshift evolution. However, the double quasar fractions in \citet{Silverman2020} and \citet{Tang2021} are {about $0.2$ dex} higher than those reported in \citet{Shen2023} and in this work. This small difference may arise from the lower secondary luminosity threshold of $L_{bol}>10^{44.3}$ erg/s adopted in \citet{Silverman2020}, while \citet{Shen2023} and our work adopt $L_{bol}>10^{45.8}$ and $L_{bol} > 10^{44.5}$ erg/s to both the primary and secondary quasars, respectively. The dual AGN fraction is generally lower for the higher-luminosity cut, as suggested by the simulations and observations \citep{Chen2023,Shen2023,Li2024}. 


In Figure \ref{fig:zevolution}, we follow Section 3.2.1 in \citet{Clara2025} and compare our results with large-scale cosmological simulations, including Illustris \citep{Genel2014,Vogelsberge2014}, TNG100, TNG300 \citep{Pillepich2018}, and Horizon-AGN \citep[HAGN;][]{Dubois2016,Volonteri2016}, using the observational constraints as presented in \citet{Silverman2020}. {Specifically, \citet{Clara2025} applies the same luminosity thresholds and further requires black hole masses of $M_{\rm BH} \geq 10^8 M_{\odot}$ and host galaxy masses of $M_* \geq 10^{10} M_{\odot}$, similar to the properties of the follow-up confirmed dual quasar systems in \citet{Silverman2020}.} Poisson errors for each simulation are included. Some simulations, such as Illustris and HAGN, do not show strong redshift evolution, whereas others do predict a declining trend (e.g., the TNG simulations show a decrease of approximately $0.5 - 1$ dex toward lower redshifts). However, this declining evolution is either absent or much weaker in the observational results \citep{Silverman2020,Shen2023}. 

In addition, simulated dual quasar fractions are typically higher than our observed values by $\sim$0.8–1.6 dex, depending on the simulation. The discrepancy can be attributed to several factors. First, \citet{Clara2025} adopt the selection criteria from \citet{Silverman2020} using a slightly brighter luminosity threshold for the secondary quasar, which may yield higher fractions than ours. Second, \citet{Clara2025} include all dual quasars in the simulations, whereas the observational studies mentioned above focus only on bright, unobscured dual quasars. Many simulations and observations suggest that many luminous dual AGNs are obscured (at least for one member) in gas-rich mergers \citep[e.g.,][]{Chen2023, Li2024}, which are not included in the observation work mentioned above leading to a higher dual AGN fraction than the observed fraction of unobscured dual quasars. Finally, although the quasar luminosities are matched between simulations and observations, there may still be deviations on the overall AGN luminosity function, and some simulations still suffer from small number statistics for the most luminous quasar population \citep[see detailed discussion in][]{Clara2025}.

Despite these differences, we conclude that the observed double quasar statistics are broadly consistent with other observations and simulation-based predictions for the dual quasar population. However, precisely assessing the consistency between simulations and observations requires future deep and wide AGN surveys. A recent work using the low-to-moderate luminosity X-ray AGN sample in the COSMOS field \citep{Li2024} showed a promising agreement between simulated and observed dual AGN fractions to fainter luminosities than probed here. But the statistics of dual AGNs in the COSMOS sample are still limited, especially at $z>3$. 

\begin{deluxetable}{cccccc}
\tablecaption{Binned Double Quasar Statistics \label{tbl:stats}}
\tablehead{
\colhead{$\Delta \theta$ (arcsec)} & \colhead{$N_{\rm QQ}$} & \colhead{$N_{\rm QQ, corr}$} & \colhead{$\sigma_{-}$} & \colhead{$\sigma_{+}$} & \colhead{$\sigma_{\rm Poisson}$}
}
\colnumbers
\startdata
0.4  & 1  & 8.0  & 8.0  & 8.0  & 8.0  \\
0.6  & 14 & 33.2 & 8.9  & 8.7  & 8.9  \\
0.8  & 12 & 17.1 & 4.7  & 4.7  & 4.9  \\
1.0  & 7  & 9.0  & 3.3  & 3.4  & 3.4  \\
1.2  & 10 & 11.3 & 3.4  & 3.4  & 3.6  \\
1.4  & 8  & 8.8  & 3.2  & 3.2  & 3.1  \\
1.6  & 4  & 4.1  & 2.0  & 2.1  & 2.0  \\
1.8  & 15 & 15.1 & 4.0  & 4.0  & 3.9  \\
2.0  & 10 & 10.0 & 3.0  & 3.0  & 3.2  \\
2.2  & 12 & 12.1 & 3.1  & 3.1  & 3.5  \\
2.4  & 12 & 12.0 & 3.0  & 3.0  & 3.5  \\
2.6  & 10 & 10.0 & 3.0  & 3.0  & 3.2  \\
2.8  & 16 & 16.1 & 4.0  & 4.0  & 4.0  \\
3.0  & 5  & 5.0  & 2.0  & 2.0  & 2.3  \\
$0.3-3.1$ & 136 & 171.9 & 8.7 & 8.7 & $\cdots$ \\
\enddata
\tablecomments{Pair statistics are measured for double quasars with $Z>0.5$ and $G<20.5$, identified through proper motion and PCA inspection, in $\Delta \theta$ bins of $0 . \! \arcsec 2$ linear size. Columns (1) lists the median separation in the unit of arcsec for each bin. Columns (2) shows the number of raw pair counts $N_{\rm QQ}$. Columns (3)–(5) are the pair statistics corrected for completeness ($N_{\rm QQ, corr}$), with the uncertainties ($\sigma_{-}$ and $\sigma_{+}$) estimated from bootstrap resampling. Column (6) lists the uncertainties in $N_{\rm QQ, corr}$ estimated from Poisson counting uncertainties from the raw pair counts $N_{\rm QQ}$.}
\end{deluxetable}

\section{Summary and Conclusions}
\label{sec:con}


In this work we performed a systematic search for galactic-scale quasar pairs combining the spectroscopic SDSS DR16 quasar catalog with data from Gaia DR3. There are $2,497$ unique Gaia pairs around SDSS quasars within a $3\arcsec$ matching radius ($\sim 3-30$ kpc at $z = 1.7$). To distinguish between true double quasars and contaminating foreground star superpositions, we apply a proper motion criterion ($\rm PMSIG < 3$), classifying companions as either quasar-like or starlike. Additionally, we perform a spectral PCA analysis for small-scale ($<1\farcs5$) pairs using SDSS spectra and conduct a literature search to further eliminate stellar contaminants. Restricting to redshift and magnitude ranges of $z > 0.5$ and $G < 20.5$ to balance sample statistics and purity, our final statistical sample consists of $136$ double quasars and $984$ quasar-star pairs. This study extends our earlier work \citep{Shen2023} to fainter quasar luminosities and lower redshifts, allowing us to examine the redshift evolution of the pair fraction with a nearly doubled sample size. We present the catalog of the full pair sample (Table~\ref{tbl:prop}) for future follow-up studies. 


With the final sample, we measure an overall double quasar fraction of approximately $5.7^{+0.3}_{-0.3} \times 10^{-4}$ within $0\farcs3 – 3\arcsec$ separations over $0.5<z\lesssim 4.5$, after correcting for pair-resolving completeness by Gaia (see Section \ref{sec:fcomp}). We find no strong redshift evolution in the dual quasar fraction over this broad redshift range. When divided at $z = 1.5$, the low-redshift subset ($\langle z \rangle = 1.0$) and the high-redshift subset ($\langle z \rangle = 2.0$) exhibit similar double quasar fractions, measured as $4.9_{-0.2}^{+0.2} \times 10^{-4}$ and $6.2_{-0.5}^{+0.5} \times 10^{-4}$, respectively.

We compare our results with previous observational studies \citep[][]{Silverman2020,Shen2023} and numerical simulations \citep[e.g.,][]{Clara2025} that focus on luminous dual quasars, and find general agreements. However, while the redshift evolution of the dual quasar fraction is generally weak in both observations and simulations, the overall dual quasar fraction in simulations is higher than the observed values. This discrepancy is likely caused by the fact that current observational searches for luminous dual quasars are largely limited to the unobscured population, missing a large fraction of dual systems where at least one quasar is obscured. 

The combination of two wide-area sky surveys, the SDSS and Gaia, offers one of the most efficient methods to identify luminous dual and lensed quasars at small separations. However, even with precise Gaia astrometric measurements to eliminate a large body of star superpositions, the confirmation of dual versus lensed quasars remains challenging, requiring dedicated, follow-up observations \citep[e.g.,][]{Chen2022d}. Most of the current dual quasar searches also exclude obscured systems, complicating a direct comparison with simulation predictions. Fortunately, upcoming wide-area space-based surveys such as Euclid \citep{Euclid} and Roman \citep{Roman} that target both obscured and unobscured AGNs at sub-arcsec resolution and extend to much fainter AGN luminosities, will become a powerhouse in discovering and characterizing the dual AGN population across cosmic history \citep{Shen2023b}. These future dual AGN samples will have greatly improved statistics and host characterization, and are much better matched to simulation predictions to study the evolution of the dual SMBH population.

\begin{acknowledgments}

We thank Clara Puerto S\'anchez and Melanie Habouzit for sharing their figure data on simulated dual quasar fractions. This work is partially supported by NSF grants AST-2108162 and AST-2206499.  
\end{acknowledgments}

\bibliography{doubleq}
\bibliographystyle{aasjournal}

\end{document}